\documentclass[12pt]{article}
\usepackage{euler}
\usepackage{amsmath}
\usepackage{amssymb}
\usepackage{amstext}
\usepackage{amscd}
\usepackage{amsfonts}
\usepackage{appendix}
\DeclareMathAlphabet{\EuFrak}{U}{euf}{m}{n}
\DeclareMathAlphabet{\EuScript}{U}{eus}{m}{n}

\title{{\bf A Solution to Non-Linear Equations of Motion of Nambu-Goto String}
\thanks{\it{This work was partially supported by Consejo
Nacional
de Investigaciones Cient\'{\i}ficas and Comisi\'{o}n de
Investigaciones Cient\'{\i}ficas de la Pcia. de Buenos
Aires;
Argentina.$\dag$Deceased}}}
\author{C.G.Bollini$^\dag$ and M.C.Rocca\\
Departamento de F\'{\i}sica, Fac. de Ciencias Exactas,\\
Universidad Nacional de La Plata.\\
C.C. 67 (1900) La Plata. Argentina.}

\date{December 15, 2009}

\begin{document}

\maketitle

\vspace{-5mm}

\begin{abstract}

In this paper we solve the non-linear Lagrange's equations
for the Nambu-Goto closed bosonic string.

We show that 
Ultradistributions of Exponential Type
(UET) are appropriate for the description
in a consistent way of string and string field theories.

We also prove that the string field 
is a linear superposition of UET of compact support (CUET),
and give the notion of anti-string.
We evaluate the propagator  for the string field,
and calculate the convolution of two of them.


\end{abstract}

\newpage

\renewcommand{\theequation}{\arabic{section}.\arabic{equation}}

\section{Introduction}

In a series of papers \cite{tp1,tp2,tp3,tp4,tp5}
we have shown that Ultradistribution theory of 
Sebastiao e Silva  \cite{tp6,tp7,tp8} permits a significant advance in the treatment 
of quantum field theory. In particular, with the use of the 
convolution of Ultradistributions we have shown that it  is possible
to define a general product of distributions ( a product in a ring
with divisors of zero) that sheds new light on  the question of the divergences
in Quantum Field Theory. Furthermore, Ultradistributions of Exponential Type  
(UET) are  adequate to describe
Gamow States and exponentially increasing fields in Quantum 
Field Theory \cite{tp9,tp10,tp11}.

In three recent papers (\cite{tq1,ts2,ts3})  we have demonstrated
that Ultradistributions of Exponential type
provide an adequate framework 
for a consistent treatment of 
string and string field theories. In particular, a general
state of the closed string is 
represented by UET of compact support,
and as a consequence the string field is a linear combination
of UET of compact support.

Ultradistributions also have the
advantage of being representable by means of analytic functions.
So that, in general, they are easier to work with  and,
as we shall see, have interesting properties. One of those properties
is that Schwartz's tempered distributions are canonical and continuously
injected into  Ultradistributions of Exponential Type
and as a consequence the Rigged
Hilbert Space  with tempered distributions is  canonical and continuously
included
in the Rigged Hilbert Space with  Ultradistributions of 
Exponential Type.

Another interesting property is that the space of UET is 
reflexive under the operation of Fourier transform (in a similar way
of tempered distributions of Schwartz)

In this paper we show that Ultradistributions of Exponential type
provides an adequate tool
for a consistent treatment of 
Nambu-Goto closed bosonic string. A general
state of the closed Nambu-Goto string is 
represented by UET of compact support,
and the corresponding string field
is a linear combination
of UET of compact support (CUET).

The motivation that inspired the writing of this paper has 
been that to quantum level the formulation of Polyakov's 
bosonic string is not equivalent to the Nambu-Goto string 
because 
$(L_0-a)|\phi>=0$, $L_m|\phi>=0$\; $m>0$ and
$L_m|\phi>\neq 0$ for $m<0$ (where $L_m$ is the
Virasoro operator and $|\phi>$ is the physical state of the string).
This implies that $T_{\alpha\beta}|\phi>\neq 0$
and then the constraints are not satisfied by the theory
because in order to satisfy $T_{\alpha\beta}|\phi>=0$
the Virasoro operators must meet $L_m|\phi>=0$
for all $m\neq 0$.
($T_{\alpha\beta}=0$ are the classical constraints of the theory).
As a consequence the solutions of the Polyakov string are not 
true solutions of the nonlinear equations of Nambu-Goto and the 
resulting theory is not equivalent to the original theory. Another 
problem presented by the Polyakov string is the presence of a 
tachyon in its ground state, whose quantification breaks
unitarity and causality of the theory.

Moreover, in his book about strings \cite{green}, Green, Schwartz and Witten
obtain in page 63
(in the proof about the equivalence of Nambu-Goto
and Polyakov string)
\begin{equation}
\label{ep1.1}
G=\frac {1} {4} h (h^{\alpha\beta}G_{\alpha\beta})^2
\end{equation}
where 
\[G=|det G_{\alpha\beta}|\]
\[h=|det h_{\alpha\beta}|\]
\[G_{\alpha\beta}={\partial}_{\alpha}X_{\mu}{\partial}_{\beta}X^{\mu}\]
and then concludes
\begin{equation}
\label{ep1.2}
\sqrt{G}=
\sqrt{h} h^{\alpha\beta}G_{\alpha\beta}
\end{equation}
and
\begin{equation}
\label{ep1.3}
\int\limits_{\Sigma}\sqrt{G}\;d^2\sigma=
\frac {1} {2} \int\limits_{\Sigma} \sqrt{h} h^{\alpha\beta}G_{\alpha\beta}
d^2\sigma
\end{equation}
In Minkowskian space
\[\int\limits_{\Sigma}\sqrt{|(\dot{X}\cdot X^{'})^2-{\dot{X}}^2 X^{'2}|}\;d^2\sigma=
\frac {1} {2} \int\limits_{\Sigma} {\dot{X}}^2-X^{'2}\;d^2\sigma\]
The right hand side of (\ref{ep1.3}) is the Polyakov action.
But this is not strictly true because
$\sqrt{G}\; d^2\sigma$ 
is the surface element of the world sheet.
Indeed we have
\begin{equation}
\label{ep1.4}
\sqrt{G}=\frac {1} {2} \sqrt{h}|h^{\alpha\beta}G_{\alpha\beta}|
\end{equation}
and then
\begin{equation}
\label{ep1.5}
\int\limits_{\Sigma}\sqrt{G}\;d^2\sigma=
\frac {1} {2} \int\limits_{\Sigma} \sqrt{h}
|h^{\alpha\beta}G_{\alpha\beta}|
d^2\sigma
\end{equation}
In Minkowskian space
\[\int\limits_{\Sigma}\sqrt{|(\dot{X}\cdot X^{'})^2-{\dot{X}}^2 X^{'2}|}\;d^2\sigma=
\frac {1} {2} \int\limits_{\Sigma} |{\dot{X}}^2-X^{'2}|\;d^2\sigma\]
(If $x$ is a real variable $+\sqrt{x^2}=|x|$)
Note then that the equations of motion  corresponding to (\ref{ep1.5})
are non-linear. 
This was the reason why we decided to solve the non-linear Nambu-Goto 
equations directly.

This paper is organized as follows:
In section 2 we solve the non-linear Lagrange's equations
for closed Nambu-Goto bosonic string.
In section 3 we give  expressions for the field of the string,
the string field propagator and the creation and annihilation
operators of a string and a anti-string.
In section 4, we give expressions for the non-local action of a free string
and a non-local interaction lagrangian for the string field similar 
to  $\lambda{\phi}^4$ in Quantum Field Theory.
Also we show how to evaluate the convolution
of two string field propagators.
In  section 5 we realize  a discussion of the principal results.
In Appendix A we define the Ultradistributions of Exponential Type
and their Fourier transform. In them we give some main results obtained
for us and other authors, used in this paper and  show that 
Ultradistributions of Exponential Type
are  part of a Guelfand's Triplet ( or Rigged Hilbert Space \cite{tp12} )
together with their respective dual and a ``middle term'' Hilbert
space. In Appendix B we give a new representation, obtained in 
\cite{tq1}, for the states 
of the string using CUET of compact support.

\section{The Closed Nambu-Goto string}

\setcounter{equation}{0}

As is known the Nambu-Goto Lagrangian for the closed bosonic string 
is given by (\cite{tp13},\cite{tp14})
\begin{equation}
\label{fp2.1}
{\cal L}_{NG}=T\sqrt{|({\dot{X}}\cdot X^{'})^{2}-{\dot{X}}^{2}X^{'2}|}
\end{equation}
where
\begin{equation}
\label{fp2.2}
\begin{cases}
X_{\mu}=X_{\mu}(\tau,\sigma)\;;\;
{\dot{X}}_{\mu}={\partial}_{\tau}X_{\mu}\;;\;X^{'}_{\mu}={\partial}_{\sigma}X_{\mu}\\
X_{\mu}(\tau,0)=X_{\mu}(\tau,\pi)\\
-\infty<\tau<\infty\;\;;\;\;0\leq\sigma\leq\pi
\end{cases}
\end{equation}
The corresponding action is:
\begin{equation}
\label{fp2.3}
{\cal S}_{NG}=T\int\limits_{-\infty}^{\infty}\int\limits_{0}^{\pi}
\sqrt{|({\dot{X}}\cdot X^{'})^{2}-{\dot{X}}^{2}X^{'2}|}\;d\sigma\;d\tau
\end{equation}
If we call
\begin{equation}
\label{fp2.4}
{\cal L}_1=({\dot{X}}\cdot X^{'})^{2}-{\dot{X}}^{2}X^{'2}
\end{equation}
The Euler-Lagrange equations are:
\[\frac {\partial} {\partial\tau} \left[Sgn({\cal L}_1)
\frac {(\dot{X}\cdot{X}^{'})X^{'}_{\mu}-X^{'2}{\dot{X}}_{\mu}} 
{\sqrt{|{\cal L}_1}|}\right]\; +\]
\begin{equation}
\label{fp2.5}
\frac {\partial} {\partial\sigma} \left[Sgn({\cal L}_1) 
\frac {(\dot{X}\cdot{X}^{'}){\dot{X}}_{\mu}-{\dot{X}}^{2}X^{'}_{\mu}} 
{\sqrt{|{\cal L}_1|}}\right]=0
\end{equation}
Let $X_{\mu}$ be given by:
\begin{equation}
\label{fp2.6}
X_{\mu}=Sgn({\dot{Y}}^2-Y^{'2}) Y_{\mu}
\end{equation}
where
\begin{equation}
\label{fp2.7}
\begin{cases}
Y_{\mu}(\tau,\sigma)=y_{\mu} + l^2 p_{\mu} \tau+\frac {il} {2}
\sum\limits_{n=-\infty\;;\;n\neq 0}^{\infty}
\frac {a_n} {n} e^{-2in(\tau-\sigma)}\\
p^2=0
\end{cases}
\end{equation}
or
\begin{equation}
\label{fp2.8}
\begin{cases}
Y_{\mu}(\tau,\sigma)=y_{\mu} + l^2p_{\mu} \tau+\frac {il} {2}
\sum\limits_{n=-\infty\;;\; n\neq 0}^{\infty}
\frac {{\tilde{a}}_n} {n} e^{-2in(\tau+\sigma)}\\
p^2=0
\end{cases}
\end{equation}
$Y_{\mu}$ of  (\ref{fp2.7}) satisfy 
\begin{equation}
\label{fp2.9}
{\dot{Y}}_{\mu}+Y^{'}_{\mu}=p_{\mu}
\end{equation}
and $Y_{\mu}$ of (\ref{fp2.8}):
\begin{equation}
\label{fp2.10}
{\dot{Y}}_{\mu}-Y^{'}_{\mu}=p_{\mu}
\end{equation}
For both we have:
\begin{equation}
\label{fp2.11} 
{\dot{X}}^2-X^{'2}={\dot{Y}}^2-Y^{'2}\neq 0
\end{equation}
and then
\begin{equation}
\label{fp2.12}
{\cal L}_1=({\dot{X}}^2-X^{'2})^2=({\dot{Y}}^2-Y^{'2})^2\neq 0
\end{equation}
We shall prove that ((\ref{fp2.6}), (\ref{fp2.7})) or ((\ref{fp2.6}),(\ref{fp2.8})) are solutions of 
(\ref{fp2.5}). From(\ref{fp2.6}), (\ref{fp2.7}) we have $\ddot{X}=-{\dot{X}}^{'}=X^{''}$
and (\ref{fp2.5}) transform into:
\begin{equation}
\label{fp2.13}
\frac {\partial} {\partial\tau} \left[
\frac {(\dot{X}\cdot{X}^{'})X^{'}_{\mu}-X^{'2}{\dot{X}}_{\mu}} 
{\sqrt{|{\cal L}_1}|} -
\frac {(\dot{X}\cdot{X}^{'}){\dot{X}}_{\mu}-{\dot{X}}^{2}X^{'}_{\mu}} 
{\sqrt{|{\cal L}_1|}}\right]=
\end{equation}
\begin{equation}
\label{fp2.14}
\frac {\partial} {\partial\tau}\left[ 
\frac {(\dot{X}\cdot{X}^{'}+{\dot{X}}^2)X^{'}_{\mu}-
(\dot{X}\cdot X^{'}+X^{'2}){\dot{X}}_{\mu}} 
{\sqrt{|{\cal L}_1}|}\right]=
\end{equation}
\begin{equation}
\label{fp2.15}
\frac {\partial} {\partial\tau}\left[ 
\frac {({\dot{X}}^2-X^{'2})X^{'}_{\mu}-
(X^{'2}-{\dot{X}}^2){\dot{X}}_{\mu}} 
{2\sqrt{|{\cal L}_1}|}\right]=
\end{equation}
\begin{equation}
\label{fp2.16}
\frac {\partial} {\partial\tau}\left[ 
\frac {({\dot{X}}^2-X^{'2})({\dot X}_{\mu}+X^{'}_{\mu})} 
{2\sqrt{|{\cal L}_1}|}\right]=
\end{equation}
and finally
\begin{equation}
\label{fp2.17}
l^2\frac {\partial p_{\mu}} {\partial\tau}=0
\end{equation}
From (\ref{fp2.6}), (\ref{fp2.8}) we have $\ddot{X}={\dot{X}}^{'}=X^{''}$ and 
(\ref{fp2.5}) transforms into:
\begin{equation}
\label{fp2.18}
\frac {\partial} {\partial\tau} \left[
\frac {(\dot{X}\cdot{X}^{'})X^{'}_{\mu}-X^{'2}{\dot{X}}_{\mu}} 
{\sqrt{|{\cal L}_1}|} +
\frac {(\dot{X}\cdot{X}^{'}){\dot{X}}_{\mu}-{\dot{X}}^{2}X^{'}_{\mu}} 
{\sqrt{|{\cal L}_1|}}\right]=
\end{equation}
\begin{equation}
\label{fp2.19}
\frac {\partial} {\partial\tau}\left[
\frac {(\dot{X}\cdot{X}^{'}-{\dot{X}}^2)X^{'}_{\mu}+
(\dot{X}\cdot X^{'}-X^{'2}){\dot{X}}_{\mu}} 
{\sqrt{|{\cal L}_1}|}\right]=
\end{equation}
\begin{equation}
\label{fp2.20}
\frac {\partial} {\partial\tau}\left[ 
\frac {({\dot{X}}^2-X^{'2}){\dot{X}}_{\mu}+
(X^{'2}-{\dot{X}}^2)X^{'}_{\mu}} 
{2\sqrt{|{\cal L}_1}|}\right]=
\end{equation}
\begin{equation}
\label{fp2.21}
\frac {\partial} {\partial\tau}\left[ 
\frac {({\dot{X}}^2-X^{'2})({\dot X}_{\mu}-X^{'}_{\mu})} 
{2\sqrt{|{\cal L}_1}|}\right]=
\end{equation}
\begin{equation}
\label{fp2.22}
l^2\frac {\partial p_{\mu}} {\partial\tau}=0
\end{equation}
At quantum level we have for (\ref{fp2.7}):
\begin{equation}
\label{fp2.23}
\begin{cases}
Y_{\mu}(\tau,\sigma)=y_{\mu} + l^2 p_{\mu} \tau+\frac {il} {2}
\sum\limits_{n=-\infty\;;\;n\neq 0}^{\infty}
\frac {a_{n\mu}} {n} e^{-2in(\tau-\sigma)}\\
p^2|\phi>=0
\end{cases}
\end{equation}
and for (\ref{fp2.8}):
\begin{equation}
\label{fp2.24}
\begin{cases}
Y_{\mu}(\tau,\sigma)=y_{\mu} + l^2p_{\mu} \tau+\frac {il} {2}
\sum\limits_{n=-\infty\;;\; n\neq 0}^{\infty}
\frac {{\tilde{a}}_{n\mu}} {n} e^{-2in(\tau+\sigma)}\\
p^2|\phi>=0
\end{cases}
\end{equation}
where $|\Phi>$ is the physical state of the string.\\
In terms of creation and annihilation operators we have:
\begin{equation}
\label{fp2.25}
\begin{cases}
Y_{\mu}(\tau,\sigma)=y_{\mu} + l^2 p_{\mu} \tau+\frac {il} {2}
\sum\limits_{n>0}
\frac {b_{n\mu}} {\sqrt{n}} e^{-2in(\tau-\sigma)}-
\frac {b^+_{n\mu}} {\sqrt{n}} e^{2in(\tau-\sigma)}\\
p^2|\phi>=0
\end{cases}
\end{equation}
\begin{equation}
\label{fp2.26}
\begin{cases}
Y_{\mu}(\tau,\sigma)=y_{\mu} + l^2p_{\mu} \tau+\frac {il} {2}
\sum\limits_{n>0}
\frac {{\tilde{b}}_{n\mu}} {\sqrt{n}} e^{-2in(\tau+\sigma)}-
\frac {{\tilde{b}}^+_{n\mu}} {\sqrt{n}} e^{-2in(\tau+\sigma)}\\
p^2|\phi>=0
\end{cases}
\end{equation}
where:
\begin{equation}
\label{fp2.27}
[b_{\mu m},b^+_{\nu n}]={\eta}_{\mu\nu}
{\delta}_{mn}
\end{equation}
\begin{equation}
\label{fp2.28}
[{\tilde{b}}_{\mu m},{\tilde{b}}^+_{\nu n}]={\eta}_{\mu\nu}
{\delta}_{mn}
\end{equation}
A general state of the string can be written as:
\[|\phi>=[a_0(p)+a^{i_1}_{\mu_1}(p)b^{+\mu_1}_{i_1}+
a^{i_1 i_2}_{\mu_1\mu_2}(p)b^{+\mu_1}_{i_1}b^{+\mu_2}_{i_2}
+...+...\]
\begin{equation}
\label{ep2.29}
+a^{i_1i_2...i_n}_{\mu_1\mu_2...\mu_n}(p)b^{+\mu_1}_{i_1}
b^{+\mu_2}_{\i_2}...b^{+\mu_n}_{i_n}+...+...]
|0>
\end{equation}
or
\[|\phi>=[a_0(p)+a^{i_1}_{\mu_1}(p){\tilde{b}}^{+\mu_1}_{i_1}+
a^{i_1 i_2}_{\mu_1\mu_2}(p){\tilde{b}}^{+\mu_1}_{i_1}{\tilde{b}}^{+\mu_2}_{i_2}
+...+...\]
\begin{equation}
\label{ep2.30}
+a^{i_1i_2...i_n}_{\mu_1\mu_2...\mu_n}(p){\tilde{b}}^{+\mu_1}_{i_1}
{\tilde{b}}^{+\mu_2}_{\i_2}...{\tilde{b}}^{+\mu_n}_{i_n}+...+...]
|0>
\end{equation}
where:
\begin{equation}
\label{ep2.31}
p^2 a^{i_1i_2...i_n}_{\mu_1\mu_2...\mu_n}(p)=0
\end{equation}

\section{The String Field}

\setcounter{equation}{0}

In this section we generalize the results of \cite{tq1} and apply these 
results to  the closed Nambu-Goto string. In this case the field of 
the string is complex. 

According to (\ref{fp2.25}), (\ref{fp2.26}) and Appendix B the equation for the string field
is given by
\begin{equation}
\label{ep8.1}
\Box\Phi(x,\{z\})=({\partial}^2_0-{\partial}^2_1-{\partial}^2_2-{\partial}^2_3)
\Phi(x,\{z\})=0
\end{equation}
where $\{z\}$ denotes $(z_{1\mu},z_{2\mu},...,z_{n\mu},...,....)$, and 
$\Phi$ is a CUET in the set of variables $\{z\}$.
Any UET of compact support can be written as a development of
$\delta(\{z\})$ and its derivatives. Thus we have:
\[\Phi(x,\{z\})=[A_0(x)+A^{i_1}_{\mu_1}(x){\partial}^{\mu_1}_{i_1}+
A^{i_1 i_2}_{\mu_1\mu_2}(x){\partial}^{\mu_1}_{i_1}{\partial}^{\mu_2}_{i_2}
+...+...\]
\begin{equation}
\label{ep8.2}
+A^{i_1i_2...i_n}_{\mu_1\mu_2...\mu_n}(x){\partial}^{\mu_1}_{i_1}
{\partial}^{\mu_2}_{\i_2}...{\partial}^{\mu_n}_{i_n}+...+...]
\delta(\{z\})
\end{equation}
where the quantum fields 
$A^{i_1i_2...i_n}_{\mu_1\mu_2...\mu_n}(x)$
are solutions of
\begin{equation}
\label{ep8.3}
\Box A^{i_1i_2...i_n}_{\mu_1\mu_2...\mu_n}(x)=0
\end{equation}
The propagator of the string field can be expressed in terms of the propagators 
of the component fields:
\[\Delta(x-x^{'},\{z\},\{z^{'}\})=[\Delta_0(x-x^{'})+\Delta^{i_1j_1}_{\mu_1\nu_1}
(x-x^{'})\partial_{i_1}^{\mu_1}\partial_{j_1}^{'\nu_1}+...+...+\]
\begin{equation}
\label{ep8.4}
\Delta^{i_1...i_nj_1...j_n}_{\mu_1...\mu_n\nu_1...\nu_n}(x-x^{'})
\partial^{\mu_1}_{i_1}...\partial^{\mu_n}_{i_n}\partial^{'\nu_1}_{j_1}...
\partial^{'\nu_n}_{j_n}+...+...]\delta(\{z\},\{z^{'}\})
\end{equation}
For the fields $A^{i_1i_2...i_n}_{\mu_1\mu_2 ...\mu_n}(x)$ we have:
\begin{equation}
\label{ep8.5}
A^{i_1i_2...i_n}_{\mu_1\mu_2...\mu_n}(x)=\int\limits_{-\infty}^{\infty}
a^{i_1i_2...i_n}_{\mu_1\mu_2...\mu_n}(k)e^{-ik_{\mu}x^{\mu}}+
b^{+i_1i_2...i_n}_{\mu_1\mu_2...\mu_n}(k)e^{ik_{\mu}x^{\mu}}\;
d^3k
\end{equation}

We define the operators of annihilation and creation of a string as:
\[a(k,\{z\})=[a_0(k)+a_{\mu_1}^{i_1}(k)\partial_{i_1}^{\mu_1}+...+...+\]
\begin{equation}
\label{ep8.6}
a_{\mu_1...\mu_n}^{i_1...i_n}(k)\partial_{i_1}^{\mu_1}...\partial_{i_n}^{\mu_n}
+...+...]\delta(\{z\})
\end{equation}
\[a^+(k^{'},\{z^{'}\})=[a^+_0(k^{'})+a_{\nu_1}^{+j_1}(k^{'})\partial_{j_1}^{'\nu_1}+...+...+\]
\begin{equation}
\label{ep8.7}
a_{\nu_1...\nu_n}^{+j_1...j_n}(k^{'})\partial_{j_1}^{'\nu_1}...\partial_{j_n}^{'\nu_n}
+...+...]\delta(\{z^{'}\})
\end{equation}
and the annihilation and creation operators for the anti-string 
\[b(k,\{z\})=[b_0(k)+b_{\mu_1}^{i_1}(k)\partial_{i_1}^{\mu_1}+...+...+\]
\begin{equation}
\label{ep8.8}
b_{\mu_1...\mu_n}^{i_1...i_n}(k)\partial_{i_1}^{\mu_1}...\partial_{i_n}^{\mu_n}
+...+...]\delta(\{z\})
\end{equation}
\[b^+(k^{'},\{z^{'}\})=[b^+_0(k^{'})+b_{\nu_1}^{+j_1}(k^{'})\partial_{j_1}^{'\nu_1}+...+...+\]
\begin{equation}
\label{ep8.9}
b_{\nu_1...\nu_n}^{+j_1...j_n}(k^{'})\partial_{j_1}^{'\nu_1}...\partial_{j_n}^{'\nu_n}
+...+...]\delta(\{z^{'}\})
\end{equation}
If we define
\begin{equation}
\label{ep8.10}
[a_{\mu_1...\mu_n}^{i_1...i_n}(k),a_{\nu_1..\nu_n}^{+j_1...j_n}(k^{'})]=
f_{\mu_1...\mu_n\nu_1...\nu_n}^{i_1...i_nj_1...j_n}(k)\delta(k-k^{'})
\end{equation}
the commutations relations are
\[[a(k,\{z\}),a^+(k^{'},\{z^{'}\})]=\delta(k-k^{'})[f_0(k)+f_{\mu_1\nu_1}^{i_1j_1}(k)
{\partial}_{i_1}^{\mu_1}{\partial}_{j_1}^{'\nu_1}+...+...\]
\begin{equation}
\label{ep8.11}
f_{\mu_1...\mu_n\nu_1...\nu_n}^{i_1...i_nj_1...j_n}(k)
{\partial}_{i_1}^{\mu_1}...{\partial}_{i_n}^{\mu_n}
{\partial}_{j_1}^{'\nu_1}...{\partial}_{j_n}^{'\nu_n}+...+...]
\delta(\{z\},\{z^{'}\})
\end{equation}
and for the anti-string:
\begin{equation}
\label{ep8.12}
[b_{\mu_1...\mu_n}^{i_1...i_n}(k),b_{\nu_1..\nu_n}^{+j_1...j_n}(k^{'})]=
g_{\mu_1...\mu_n\nu_1...\nu_n}^{i_1...i_nj_1...j_n}(k)\delta(k-k^{'})
\end{equation}
the commutations relations are
\[[b(k,\{z\}),b^+(k^{'},\{z^{'}\})]=\delta(k-k^{'})[g_0(k)+g_{\mu_1\nu_1}^{i_1j_1}(k)
{\partial}_{i_1}^{\mu_1}{\partial}_{j_1}^{'\nu_1}+...+...\]
\begin{equation}
\label{ep8.13}
g_{\mu_1...\mu_n\nu_1...\nu_n}^{i_1...i_nj_1...j_n}(k)
{\partial}_{i_1}^{\mu_1}...{\partial}_{i_n}^{\mu_n}
{\partial}_{j_1}^{'\nu_1}...{\partial}_{j_n}^{'\nu_n}+...+...]
\delta(\{z\},\{z^{'}\})
\end{equation}
With this annihilation and creation operators we can write:
\begin{equation}
\label{ep8.14}
\Phi(x,\{z\})=\int\limits_{-\infty}^{\infty}
a(k,\{z\})e^{-ik_{\mu}x^{\mu}}+
b^+(k\{z\})e^{ik_{\mu}x^{\mu}}\;
d^3k
\end{equation}

\section{The Action for the String Field}

\setcounter{equation}{0}

\subsection*{The case n finite}

In this section we generalize the results of \cite{tq1} for a
complex string field

The action for the free bosonic closed string field is:
\begin{equation}
\label{ep9.1}
S_{free}=
\oint\limits_{\{\Gamma_1\}}\oint\limits_{\{\Gamma_2\}}
 \int\limits_{-\infty}^{\infty}
\partial_{\mu}\Phi(x,\{z_1\})e^{\{z_1\}{\cdot}\{z_2\}}
\partial^{\mu}\Phi^+(x,\{z_2\})\;d^3x\;\{dz_1\}\;\{dz_2\}
\end{equation}
A possible interaction is given by:
\[S_{int}=\lambda\;\oint\limits_{\{\Gamma_1\}}
\oint\limits_{\{\Gamma_2\}}
\oint\limits_{\{\Gamma_3\}}
\oint\limits_{\{\Gamma_4\}}
\int\limits_{-\infty}^{\infty}
\Phi(x,\{z_1\})e^{\{z_1\}{\cdot}\{z_2\}}
\Phi^+(x,\{z_2\})e^{\{z_2\}{\cdot}\{z_3\}}
\Phi(x,\{z_3\})\times\]
\begin{equation}
\label{ep9.2}
e^{\{z_3\}{\cdot}\{z_4\}}
\Phi^+(x,\{z_4\})\;
d^3x\;\{dz_1\}\;\{dz_2\}
\;\{dz_3\}\;\{dz_4\}
\end{equation}
Both, $S_{free}$ and $S_{int}$ are non-local as expected.

\subsection*{The case n$\rightarrow \infty$}

In this case:
\begin{equation}
\label{ep9.3}
[S_{free}=
\oint\limits_{\{\Gamma_1\}}\oint\limits_{\{\Gamma_2\}}
 \int\limits_{-\infty}^{\infty}
\partial_{\mu}\Phi(x,\{z_1\})
e^{\{z_1\}{\cdot}\{z_2\}}
\partial^{\mu}\Phi^+(x,\{z_2\})\;d^3x\;\{d\eta_1\}\;\{d\eta_2\}
\end{equation}
where
\begin{equation}
\label{ep9.4}
d\eta_{i\mu}=\frac {e^{-z_{i\mu}^2}} {\sqrt{\sqrt{2}\;\pi}} dz_{i\mu}
\end{equation}
and
\[S_{int}=\lambda\;\oint\limits_{\{\Gamma_1\}}
\oint\limits_{\{\Gamma_2\}}
\oint\limits_{\{\Gamma_3\}}
\oint\limits_{\{\Gamma_4\}}
\int\limits_{-\infty}^{\infty}
\Phi(x,\{z_1\})e^{\{z_1\}{\cdot}\{z_2\}}
\Phi^+(x,\{z_2\})e^{\{z_2\}{\cdot}\{z_3\}}
\Phi(x,\{z_3\})\times\]
\begin{equation}
\label{ep9.5}
e^{\{z_3\}{\cdot}\{z_4\}}
\Phi^+(x,\{z_4\})\;
d^3x\;\{d\eta_1\}\;\{d\eta_2\}
\;\{d\eta_3\}\;\{d\eta_4\}
\end{equation}

\subsection*{Gauge Conditions}

The gauge conditions for the string field are:
\begin{equation}
\label{ep9.6}
\int\limits_{\{\Gamma\}}z_{i_1}^{\mu_1}\cdot\cdot\cdot
z_{i_k}^{\mu_k} {\partial}_{\mu_k}\cdot\cdot\cdot z_{i_n}^{\mu_n}
\Phi(x,\{z\})\;\{dz\}=0
\end{equation}
${\partial}_{{\mu}_k}=\partial / \partial x^{\mu_k}\;;\;1\leq k\leq n\;;\;n\geq 1$ \\
With these gauge conditions the number of the components fields 
of the string field is finite, and the temporal components of all
fields are eliminated.

Another gauge conditions that can be added to (\ref{ep9.6}) are
\begin{equation}
\label{ep9.7}
\int\limits_{\{\Gamma\}}z_{i_1}^{\mu_1}\cdot\cdot\cdot
z_{i_k}^{\mu_k}\cdot\cdot\cdot z_{i_n}^{\mu_n}
\Phi(x,\{z\})\;\{dz\}=0\;\;\;;\;\;\;1\leq k\leq n\;\;\;;\;\;\;n\geq 1
\end{equation}
$1\leq k\leq n\;;\;n\geq 1$\\
These additional gauge conditions permit us nullify other 
component fields according to experimental data.
It should be noted that gauge conditions (\ref{ep9.6}) and
(\ref{ep9.7})  does not modify the equations of motion of
string field.

The convolution
of two propagators of  the string field is:
\begin{equation}
\label{ep9.8}
\hat{\Delta}(k,\{z_1\},\{z_2\})\ast
\hat{\Delta}(k,\{z_3\},\{z_4\})
\end{equation}
where $\ast$ denotes the convolution
of Ultradistributions of Exponential Type  
on the $k$ variable only.
With the use of the result
\begin{equation}
\label{ep9.9}
\frac {1} {\rho}\ast\frac {1} {\rho}=-\pi^2\ln\rho
\end{equation}
($\rho=x_0^2+x_1^2+x_2^2+x_3^2$ in euclidean space)

and
\begin{equation}
\label{ep9.10}
\frac {1} {\rho\pm i0}\ast\frac {1} {\rho\pm i0}=
 \mp i\pi^2\ln(\rho\pm i0)
\end{equation}
($\rho=x_0^2-x_1^2-x_2^2-x_3^2$ in minkowskian space)

the convolution of two string field propagators is finite.

\section{Discussion}

In this paper
we have shown that UET are appropriate for
the description 
in a consistent way of string and string field theories.
We have solved the non-linear Lagrange's equations corresponding 
to Nambu-Goto Lagrangian. Also we have obtained the equations of 
motion  for the field of the string and solve it with the use of CUET.
We have proved that this string field 
is a linear superposition of CUET.
We have evaluated the propagator for the string field,
and calculate the convolution of two of them, taking
into account that string field theory is a non-local theory
of  UET of an infinite number of complex variables,
For practical calculations and experimental results 
we have given expressions that
involve only a finite number of variables.
 
We have decided to include, for the benefit
of the reader, a first appendix with a summary of the main characteristics
of Ultradistributions of Exponential Type and their Fourier
transform used in this paper and a second appendix with
the representation of the states of the closed string obtained in \cite{tq1}.

As a final remark we would like to point out that our formulas
for convolutions follow from  general definitions. They are not
regularized expressions.

\newpage

\renewcommand{\thesection}{\Alph{section}}

\renewcommand{\theequation}{\Alph{section}.\arabic{equation}}

\setcounter{section}{1}

\section*{Appendix A}

\setcounter{equation}{0}

\subsection*{Ultradistributions of Exponential Type}

Let ${\cal S}$ be the Schwartz space of rapidly decreasing test functions. 
Let ${\Lambda}_j$ be the region of the complex plane defined as:
\begin{equation}
\label{er2.1}
{\Lambda}_j=\left\{z\in\boldsymbol{\mathbb{C}} :
|\Im(z)|< j : j\in\boldsymbol{\mathbb{N}}\right\}
\end{equation}
According to ref.\cite{tp6,tp8} be the space of test functions $\hat{\phi}\in
{\large{V}}_j$ is
constituted by all entire analytic functions of ${\cal S}$ for which
\begin{equation}
\label{ep2.2}
||\hat{\phi} ||_j=\max_{k\leq j}\left\{\sup_{z\in{\Lambda}_j}\left[e^{(j|\Re (z)|)}
|{\hat{\phi}}^{(k)}(z)|\right]\right\}
\end{equation}
is finite.\\
The space ${\large{Z}}$ is then defined as:
\begin{equation}
\label{er2.3}
\large{Z} =\bigcap_{j=0}^{\infty} {\large{V}}_j
\end{equation}
It is a complete countably normed space with the topology generated by
the system of semi-norms $\{||\cdot ||_j\}_{j\in \mathbb{N}}$.
The dual of $\large{Z}$, denoted by
$\large{B}$, is by definition the space of ultradistributions of exponential
type (ref.\cite{tp6,tp8}).
Let $S$ the space of rapidly decreasing sequences. According to
ref.\cite{tp12} $S$ is a nuclear space. We consider now the space of
sequences $P$ generated by the Taylor development of
$\hat{\phi}\in\large{Z}$
\begin{equation}
\label{er2.4}
P=\left\{Q : Q
\left(\hat{\phi}(0),{\hat{\phi}}^{'}(0),\frac {{\hat{\phi}}^{''}(0)} {2},...,
\frac {{\hat{\phi}}^{(n)}(0)} {n!},...\right) : \hat{\phi}\in Z\right\}
\end{equation}
The norms that define the topology of $P$ are given by:
\begin{equation}
\label{er2.5}
||\hat{\phi} ||^{'}_p=\sup_n \frac {n^p} {n} |{\hat{\phi}}^n(0)|
\end{equation}
$P$ is a subspace of $S$ and therefore is a nuclear space.
As the norms $||\cdot ||_j$ and $||\cdot ||^{'}_p$ are equivalent, the correspondence
\begin{equation}
\label{er2.6}
\large{Z}\Longleftrightarrow P
\end{equation}
is an isomorphism and therefore $Z$ is a countably normed nuclear space.
We can define now the set of scalar products
\[<\hat{\phi}(z),\hat{\psi}(z)>_n=\sum\limits_{q=0}^n\int\limits_{-\infty}^{\infty}e^{2n|z|}
\overline{{\hat{\phi}}^{(q)}}(z){\hat{\psi}}^{(q)}(z)\;dz=\]
\begin{equation}
\label{er2.7}
\sum\limits_{q=0}^n\int\limits_{-\infty}^{\infty}e^{2n|x|}
\overline{{\hat{\phi}}^{(q)}}(x){\hat{\psi}}^{(q)}(x)\;dx
\end{equation}
This scalar product induces the norm
\begin{equation}
\label{er2.8}
||\hat{\phi}||_n^{''}=[<\hat{\phi}(x),\hat{\phi}(x)>_n]^{\frac {1} {2}}
\end{equation}
The norms $||\cdot ||_j$ and $||\cdot ||^{''}_n$ are equivalent, and therefore
$\large{Z}$ is a countably hilbertian nuclear space.
Thus, if we call now ${\large{Z}}_p$ the completion of
$\large{Z}$ by the norm $p$ given in (\ref{er2.8}), we have:
\begin{equation}
\label{er2.9}
\large{Z}=\bigcap_{p=0}^{\infty}{\large{Z}}_p
\end{equation}
where
\begin{equation}
\label{er2.10}
{\large{Z}}_0=\boldsymbol{H}
\end{equation}
is the Hilbert space of square integrable functions.\\
As a consequence the ``nested space''
\begin{equation}
\label{er2.11}
\Large{U}=\boldsymbol{(}\large{Z},
\boldsymbol{H}, \large{B}\boldsymbol{)}
\end{equation}
is a Guelfand's triplet (or a Rigged Hilbert space=RHS. See ref.\cite{tp12}).

Any Guelfand's triplet
$\Large{G}=\boldsymbol{(}\boldsymbol{\Phi},
\boldsymbol{H},\boldsymbol{{\Phi}^{'}}\boldsymbol{)}$
has the fundamental property that a linear and symmetric operator
on $\boldsymbol{\Phi}$, admitting an extension to a self-adjoint
operator in
$\boldsymbol{H}$, has a complete set of generalized eigen-functions
in $\boldsymbol{{\Phi}^{'}}$ with real eigenvalues.

$\large{B}$ can also be characterized in the following way
( refs.\cite{tp6},\cite{tp8} ): let ${E}_{\omega}$ be the space of
all functions $\hat{F}(z)$ such that:

${\Large {\boldsymbol{I}}}$-
$\hat{F}(z)$ is analytic for $\{z\in \boldsymbol{\mathbb{C}} :
|Im(z)|>p\}$.

${\Large {\boldsymbol{II}}}$-
$\hat{F}(z)e^{-p|\Re(z)|}/z^p$ is bounded continuous  in
$\{z\in \boldsymbol{\mathbb{C}} :|Im(z)|\geqq p\}$,
where $p=0,1,2,...$ depends on $\hat{F}(z)$.

Let $N$ be:
$N=\{\hat{F}(z)\in{E}_{\omega} :\hat{F}(z)\; \rm{is\; entire\; analytic}\}$.
Then $\large{B}$ is the quotient space:

${\Large {\boldsymbol{III}}}$-
$\large{B}={E}_{\omega}/N$

Due to these properties it is possible to represent any ultradistribution
as ( ref.\cite{tp6,tp8} ):
\begin{equation}
\label{er2.12}
\hat{F}(\hat{\phi})=<\hat{F}(z), \hat{\phi}(z)>=\oint\limits_{\Gamma} \hat{F}(z) \hat{\phi}(z)\;dz
\end{equation}
where the path ${\Gamma}_j$ runs parallel to the real axis from
$-\infty$ to $\infty$ for $Im(z)>\zeta$, $\zeta>p$ and back from
$\infty$ to $-\infty$ for $Im(z)<-\zeta$, $-\zeta<-p$.
( $\Gamma$ surrounds all the singularities of $\hat{F}(z)$ ).

Formula (\ref{er2.12}) will be our fundamental representation for a tempered
ultradistribution. Sometimes use will be made of ``Dirac formula''
for exponential ultradistributions ( ref.\cite{tp6} ):
\begin{equation}
\label{er2.13}
\hat{F}(z)\equiv\frac {1} {2\pi i}\int\limits_{-\infty}^{\infty}
\frac {\hat{f}(t)} {t-z}\;dt\equiv
\frac {\cosh(\lambda z)} {2\pi i}\int\limits_{-\infty}^{\infty}
\frac {\hat{f}(t)} {(t-z)\cosh(\lambda t)}\;dt
\end{equation}
where the ``density'' $\hat{f}(t)$ is such that
\begin{equation}
\label{er2.14}
\oint\limits_{\Gamma} \hat{F}(z) \hat{\phi}(z)\;dz =
\int\limits_{-\infty}^{\infty} \hat{f}(t) \hat{\phi}(t)\;dt
\end{equation}
(\ref{er2.13}) should be used carefully.
While $\hat{F}(z)$ is analytic on $\Gamma$, the density $\hat{f}(t)$ is in
general singular, so that the r.h.s. of (\ref{er2.14}) should be interpreted
in the sense of distribution theory.

Another important property of the analytic representation is the fact
that on $\Gamma$, $\hat{F}(z)$ is bounded by a exponential and a power of $z$
( ref.\cite{tp6,tp8} ):
\begin{equation}
\label{er2.15}
|\hat{F}(z)|\leq C|z|^pe^{p|\Re(z)|}
\end{equation}
where $C$ and $p$ depend on $\hat{F}$.

The representation (\ref{er2.12}) implies that the addition of any entire function
$\hat{G}(z)\in N$ to $\hat{F}(z)$ does not alter the ultradistribution:
\[\oint\limits_{\Gamma}\{\hat{F}(z)+\hat{G}(z)\}\hat{\phi}(z)\;dz=
\oint\limits_{\Gamma} \hat{F}(z)\hat{\phi}(z)\;dz+\oint\limits_{\Gamma}
\hat{G}(z)\hat{\phi}(z)\;dz\]
But:
\[\oint\limits_{\Gamma} \hat{G}(z)\hat{\phi}(z)\;dz=0\]
as $\hat{G}(z)\hat{\phi}(z)$ is entire analytic
( and rapidly decreasing ),
\begin{equation}
\label{er2.16}
\therefore \;\;\;\;\oint\limits_{\Gamma} \{\hat{F}(z)+\hat{G}(z)\}\hat{\phi}(z)\;dz=
\oint\limits_{\Gamma} \hat{F}(z)\hat{\phi}(z)\;dz
\end{equation}

Another very important property of $\large{B}$ is that
$\large{B}$ is reflexive under the Fourier transform:
\begin{equation}
\label{er2.17}
\large{B}={\cal F}_c\left\{\large{B}\right\}=
{\cal F}\left\{\large{B}\right\}
\end{equation}
where the complex Fourier transform $F(k)$ of $\hat{F}(z)\in\large{B}$
is given by:
\[F(k)=\Theta[\Im(k)]\int\limits_{{\Gamma}_+}\hat{F}(z)e^{ikz}\;dz-
\Theta[-\Im(k)]\int\limits_{{\Gamma}_{-}}\hat{F}(z)e^{ikz}\;dz=\]
\begin{equation}
\label{er2.18}
\Theta[\Im(k)]\int\limits_0^{\infty}\hat{f}(x)e^{ikx}\;dx-
\Theta[-\Im(k)]\int\limits_{-\infty}^0\hat{f}(x) e^{ikx}\;dx
\end{equation}
Here ${\Gamma}_+$ is the part of $\Gamma$ with $\Re(z)\geq 0$ and
${\Gamma}_{-}$ is the part of $\Gamma$ with $\Re(z)\leq 0$
Using (\ref{er2.18}) we can interpret Dirac's formula as:
\begin{equation}
\label{er2.19}
F(k)\equiv\frac {1} {2\pi i}\int\limits_{-\infty}^{\infty}
\frac {f(s)} {s-k}\; ds\equiv{\cal F}_c\left\{{\cal F}^{-1}\left\{f(s)\right\}\right\}
\end{equation}
The treatment for ultradistributions of exponential type defined on
${\boldsymbol{\mathbb{C}}}^n$ is similar to the case of one variable.
Thus
\begin{equation}
\label{er2.20}
{\Lambda}_j=\left\{z=(z_1, z_2,...,z_n)\in{\boldsymbol{\mathbb{C}}}^n :
|\Im(z_k)|\leq j\;\;\;1\leq k\leq n\right\}
\end{equation}
\begin{equation}
\label{er2.21}
||\hat{\phi} ||_j=\max_{k\leq j}\left\{\sup_{z\in{\Lambda}_j}\left[
e^{j\left[\sum\limits_{p=1}^n|\Re(z_p)|\right]}\left| D^{(k)}\hat{\phi}(z)\right|\right]\right\}
\end{equation}
where $D^{(k)}={\partial}^{(k_1)}{\partial}^{(k_2)}\cdot\cdot\cdot{\partial}^{(k_n)}\;\;\;\;
k=k_1+k_2+\cdot\cdot\cdot+k_n$

${\large{B}}^n$ is characterized as follows. Let
${E}^n_{\omega}$ be the space of all functions $\hat{F}(z)$ such that:

${\Large {\boldsymbol{I}}}^{'}$-
$\hat{F}(z)$ is analytic for $\{z\in \boldsymbol{{\mathbb{C}}^n} :
|Im(z_1)|>p, |Im(z_2)|>p,...,|Im(z_n)|>p\}$.

${\Large {\boldsymbol{II}}}^{'}$-
$\hat{F}(z)e^{-\left[p\sum\limits_{j=1}^n|\Re(z_j)|\right]}/z^p$
is bounded continuous  in
$\{z\in \boldsymbol{{\mathbb{C}}^n} :|Im(z_1)|\geqq p,|Im(z_2)|\geqq p,
...,|Im(z_n)|\geqq p\}$,
where $p=0,1,2,...$ depends on $\hat{F}(z)$.

Let ${N}^n$ be:
${N}^n=\left\{\hat{F}(z)\in{E}^n_{\omega} :\hat{F}(z)\;\right.$
is entire analytic at minus in one of the variables $\left. z_j\;\;\;1\leq j\leq n\right\}$
Then ${\large{B}}^n$ is the quotient space:

${\Large {\boldsymbol{III}}}^{'}$-
${\large{B}}^n={E}^n_{\omega}/{N}^n$
We have now
\begin{equation}
\label{er2.22}
\hat{F}(\hat{\phi})=<\hat{F}(z), \hat{\phi}(z)>=\oint\limits_{\Gamma} \hat{F}(z) \hat{\phi}(z)\;
dz_1\;dz_2\cdot\cdot\cdot dz_n
\end{equation}
$\Gamma={\Gamma}_1\cup{\Gamma}_2\cup ...{\Gamma}_n$
where the path ${\Gamma}_j$ runs parallel to the real axis from
$-\infty$ to $\infty$ for $Im(z_j)>\zeta$, $\zeta>p$ and back from
$\infty$ to $-\infty$ for $Im(z_j)<-\zeta$, $-\zeta<-p$.
(Again $\Gamma$ surrounds all the singularities of $\hat{F}(z)$ ).
The n-dimensional Dirac's formula is
\begin{equation}
\label{ep2.23}
\hat{F}(z)=\frac {1} {(2\pi i)^n}\int\limits_{-\infty}^{\infty}
\frac {\hat{f}(t)} {(t_1-z_1)(t_2-z_2)...(t_n-z_n)}\;dt_1\;dt_2\cdot\cdot\cdot dt_n
\end{equation}
where the ``density'' $\hat{f}(t)$ is such that
\begin{equation}
\label{ep2.24}
\oint\limits_{\Gamma} \hat{F}(z)\hat{\phi}(z)\;dz_1\;dz_2\cdot\cdot\cdot dz_n =
\int\limits_{-\infty}^{\infty} f(t) \hat{\phi}(t)\;dt_1\;dt_2\cdot\cdot\cdot dt_n
\end{equation}
and the modulus of $\hat{F}(z)$ is bounded by
\begin{equation}
\label{er2.25}
|\hat{F}(z)|\leq C|z|^p e^{\left[p\sum\limits_{j=1}^n|\Re(z_j)|\right]}
\end{equation}
where $C$ and $p$ depend on $\hat{F}$.

\subsection{The Case N$\rightarrow\infty$}

When the number of variables of the argument of the Ultradistribution of 
Exponential type tends to infinity we define:
\begin{equation}
\label{ep3.1}
d\mu(x)=\frac {e^{-x^2}} {\sqrt{\pi}}dx
\end{equation}
Let $\hat{\phi}(x_1,x_2,...,x_n)$ be such that:
\begin{equation}
\label{ep3.2}
\idotsint\limits_{-\infty}^{\;\;\infty}|\hat{\phi}(x_1,x_2,...,x_n)|^2 d{\mu}_1d{\mu}_2...
d{\mu}_n<\infty
\end{equation}
where
\begin{equation}
\label{ep3.3}
d{\mu}_i=\frac {e^{-x_i^2}} {\sqrt{\pi}}dx_i
\end{equation}
Then by definition
$\hat{\phi}(x_1,x_2,...,x_n)\in L_2({\mathbb{R}}^n,\mu)$
and 
\begin{equation}
\label{ep3.4}
L_2({\mathbb{R}}^{\infty},\mu)=
\bigcup\limits_{n=1}^{\infty}L_2({\mathbb{R}}^n,\mu)
\end{equation}
Let $\hat{\psi}$ be given by
\begin{equation}
\label{ep3.5}
\hat{\psi}(z_1,z_2,...,z_n)={\pi}^{n/4}\hat{\phi}(z_1,z_2,...,z_n)
e^{\frac {z_1^2+z_2^2+...+z_n^2} {2}}
\end{equation}
where $\hat{\phi}\in {\large{Z}}^n$(the corresponding 
n-dimensional of $\large{Z}$).\\
Then by definition $\hat{\psi}(z_1,z_2,...,z_n)\in \large{G}({\mathbb{C}}^n)$,
\begin{equation}
\label{ep3.6}
\large{G}({\mathbb{C}}^{\infty})=\bigcup\limits_{n=1}^{\infty}
\large{G}({\mathbb{C}}^n)
\end{equation}
and the dual ${\large{G}}^{'}({\mathbb{C}}^{\infty})$ given by
\begin{equation}
\label{ep3.7}
{\large{G}}^{'}({\mathbb{C}}^{\infty})=\bigcup\limits_{n=1}^{\infty}
{\large{G}}^{'}({\mathbb{C}}^n)
\end{equation}
is the space of Ultradistributions of Exponential type.\\
The analog to (\ref{er2.11}) in the infinite dimensional case is:
\begin{equation}
\label{ep3.8}
{\Large{W}}=\boldsymbol{(}{\large{G}}({\mathbb{C}}^{\infty}),
L_2({\mathbb{R}}^{\infty},\mu), 
{\large{G}}^{'}({\mathbb{C}}^{\infty})\boldsymbol{)}
\end{equation}
If we define:
\begin{equation}
\label{ep3.9}
{\cal F}:{\large{G}}({\mathbb{C}}^{\infty})\rightarrow
{\large{G}}({\mathbb{C}}^{\infty})
\end{equation}
via the Fourier transform:
\begin{equation}
\label{ep3.10}
{\cal F}:{\large{G}}({\mathbb{C}}^n)\rightarrow
{\large{G}}({\mathbb{C}}^n)
\end{equation}
given by:
\begin{equation}
\label{ep3.11}
{\cal F}\{\hat{\psi}\}(k)=
\int\limits_{-\infty}^{\infty}\hat{\psi}(z_1,z_2,...,z_n)
e^{ik\cdot z+\frac {k^2} {2}}d{\rho}_1d{\rho}_2...d{\rho}_n
\end{equation}
where
\begin{equation}
\label{ep3.12}
d\rho(z)=\frac {e^{-\frac {z^2} {2}}} {\sqrt{2\pi}}\;dz
\end{equation}
we conclude that
\begin{equation}
\label{ep3.13}
{\large{G}}^{'}({\mathbb{C}}^{\infty})=
{\cal F}_c\{{\large{G}}^{'}({\mathbb{C}}^{\infty})\}=
{\cal F}\{{\large{G}}^{'}({\mathbb{C}}^{\infty})\}
\end{equation}
where in the one-dimensional case
\begin{equation}
\label{ep3.14}
{\cal F}_c\{\hat{\psi}\}(k)=
\Theta[\Im(k)]\int\limits_{{\Gamma}_+}\hat{\psi}(z)e^{ikz+\frac {k^2} {2}}\;d\rho-
\Theta[-\Im(k)]\int\limits_{{\Gamma}_{-}}\hat{\psi}(z)e^{ikz+\frac {k^2} {2}}\;d\rho
\end{equation}

\setcounter{section}{2}

\section*{Appendix B} 

\subsection*{A representation of  the states of the Closed String}

\setcounter{equation}{0}

\subsection*{The case n finite}

For an ultradistribution of exponential type, we can write:
\[G(k)=\oint\limits_{{\Gamma}_z}\left\{\Theta[\Im(k)]\Theta[\Re(z)]-
\Theta[-\Im(k)]\Theta[-\Re(z)]\right\}\hat{G}(z)
e^{ikz}\;dz\]
\begin{equation}
\label{ep7.1}
\hat{G}(z)=\frac {1} {2\pi}\oint\limits_{{\Gamma}_k}\left\{\Theta[\Im(z)]\Theta[-\Re(k)]-
\Theta[-\Im(z)]\Theta[\Re(k)]\right\}G(k)
e^{-ikz}\;dk
\end{equation}
and 
\[G(\phi)=\oint\limits_{{\Gamma}_k}
G(k)\phi(k)\;dk=\]
\begin{equation}
\label{ep7.2}
\oint\limits_{{\Gamma}_k}\oint\limits_{{\Gamma}_z}
\left\{\Theta[\Im(k)]\Theta[\Re(z)]-
\Theta[-\Im(k)]\Theta[-\Re(z)]\right\}\hat{G}(z)
\phi(k)e^{ikz}\;dk\;dz=
\end{equation}
\begin{equation}
\label{ep7.3}
-i\oint\limits_{{\Gamma}_k}\oint\limits_{{\Gamma}^{'}_z}
\left\{\Theta[\Im(k)]\Theta[\Im(z)]-
\Theta[-\Im(k)]\Theta[-\Im(z)]\right\}\hat{G}(-iz)
\phi(k)e^{kz}\;dk\;dz
\end{equation}
where the path ${\Gamma}^{'}_z$ is the path ${\Gamma}_z$
rotated ninety degrees counterclockwise around the origin
of the complex plane.

If $F(z)$ is an UET of compact support we can define: 
\[<\hat{F}(z),\phi(z)>=\]
\begin{equation}
\label{ep7.4}
\oint\limits_{{\Gamma}_k}\oint\limits_{{\Gamma}^{'}_z}
\left\{\Theta[\Im(k)]\Theta[\Im(z)]-
\Theta[-\Im(k)]\Theta[-\Im(z)]\right\}\hat{F}(z)
\phi(k)e^{kz}\;dk\;dz
\end{equation}
then:
\[<{\hat{F}}^{'}(z),\phi(z)>=\]
\[\oint\limits_{{\Gamma}_k}\oint\limits_{{\Gamma}^{'}_z}
\left\{\Theta[\Im(k)]\Theta[\Im(z)]-
\Theta[-\Im(k)]\Theta[-\Im(z)]\right\}{\hat{F}}^{'}(z)
\phi(k)e^{kz}\;dk\;dz=\]
\[-\oint\limits_{{\Gamma}_k}\oint\limits_{{\Gamma}^{'}_z}
\left\{\Theta[\Im(k)]\Theta[\Im(z)]-
\Theta[-\Im(k)]\Theta[-\Im(z)]\right\}{\hat{F}}(z)
k\phi(k)e^{kz}\;dk\;dz=\]
\begin{equation}
\label{ep7.5}
<\hat{F}(z),-z\phi(z)>
\end{equation}
If we define:
\begin{equation}
\label{ep7.6}
a=-z\;\;\;;\;\;\;a^+=\frac {d} {dz}
\end{equation}
we have
\begin{equation}
\label{ep7.7}
[a,a^+]=I
\end{equation}
Thus we have a representation for creation and annihilation operators
of the states of the string. The vacuum state annihilated 
by $z_{\mu}$ is the UET $\delta(z_{\mu})$,
and the orthonormalized states obtained by successive application of 
$\frac {d} {dz_{\mu}}$ to $\delta(z_{\mu})$ are:
\begin{equation}
\label{ep7.8}
F_n(z_{\mu})=\frac {{\delta}^{(n)}(z_{\mu})} {\sqrt{n!}}
\end{equation}
On the real axis:
\begin{equation}
\label{ep7.9}
<\hat{F}(z),\phi(z)>=\int\limits_{-\infty}^{\infty}\int\limits_{-\infty}^{\infty}
\overline{\hat{f}}(x)\phi(k)e^{kx}\;dx\;dk
\end{equation}
where $\overline{\hat{f}}(x)$is given by Dirac's formula:
\begin{equation}
\label{ep7.10}
\hat{F}(z)=\frac {1} {2\pi i}\int\limits_{-\infty}^{\infty}
\frac {\overline{\hat{f}}(x)} {x-z}\;dx
\end{equation}

A general state of the string can be written as:
\[\phi(x,\{z\})=[a_0(x)+a^{i_1}_{\mu_1}(x){\partial}^{\mu_1}_{i_1}+
a^{i_1 i_2}_{\mu_1\mu_2}(x){\partial}^{\mu_1}_{i_1}{\partial}^{\mu_2}_{i_2}
+...+...\]
\begin{equation}
\label{ep7.11}
+a^{i_1i_2...i_n}_{\mu_1\mu_2...\mu_n}(x){\partial}^{\mu_1}_{i_1}
{\partial}^{\mu_2}_{\i_2}...{\partial}^{\mu_n}_{i_n}+...+...]
\delta(\{z\})
\end{equation}
where $\{z\}$ denotes $(z_{1\mu},z_{2\mu},...,z_{n\mu},...,....)$, and 
$\phi$ is a UET of compact support in the set of variables $\{z\}$.
The functions
$a^{i_1i_2...i_n}_{\mu_1\mu_2...\mu_n}(x)$
are solutions of
\begin{equation}
\label{ep7.12}
\Box a^{i_1i_2...i_n}_{\mu_1\mu_2...\mu_n}(x)=0
\end{equation}

\subsection*{The case n$\rightarrow \infty$}

In this case
\[G(k)=\oint\limits_{{\Gamma}_z}\left\{\Theta[\Im(k)]\Theta[\Re(z)]-
\Theta[-\Im(k)]\Theta[-\Re(z)]\right\}\hat{G}(z)
e^{ikz+\frac {k^2} {2}-\frac {z^2} {2}}\;\frac {dz} {\sqrt{2\pi}}\]
\[\hat{G}(z)=\oint\limits_{{\Gamma}_k}\left\{\Theta[\Im(z)]\Theta[-\Re(k)]-
\Theta[-\Im(z)]\Theta[\Re(k)]\right\}\times\]
\begin{equation}
\label{ep7.13}
G(k)
e^{-ikz+\frac {z^2} {2}-\frac {k^2} {2}}\;\frac {dk} {\sqrt{2\pi}}
\end{equation}
\[G(\phi)=\oint\limits_{{\Gamma}_k}
G(k)\phi(k)e^{-k^2}\;\frac {dk} {\sqrt{\pi}}=\]
\[\oint\limits_{{\Gamma}_k}\oint\limits_{{\Gamma}_z}
\left\{\Theta[\Im(k)]\Theta[\Re(z)]-
\Theta[-\Im(k)]\Theta[-\Re(z)]\right\}\times\]
\begin{equation}
\label{ep7.14}
\hat{G}(z)
\phi(k)e^{ikz-\frac {z^2} {2}-k^2}\;\frac {dk\;dz} {\sqrt{2}\;\pi}=
\end{equation}
\[-i\oint\limits_{{\Gamma}_k}\oint\limits_{{\Gamma}^{'}_z}
\left\{\Theta[\Im(k)]\Theta[\Im(z)]-
\Theta[-\Im(k)]\Theta[-\Im(z)]\right\}\times\]
\begin{equation}
\label{ep7.15}
\hat{G}(-iz)
\phi(k)e^{kz+\frac {z^2} {2}-k^2}\;\frac {dk\;dz} {\sqrt{2}\;\pi}
\end{equation}
If $F(z)$ is an CUET we can define: 
\[<\hat{F}(z),\phi(z)>=\]
\[\oint\limits_{{\Gamma}_k}\oint\limits_{{\Gamma}^{'}_z}
\left\{\Theta[\Im(k)]\Theta[\Im(z)]-
\Theta[-\Im(k)]\Theta[-\Im(z)]\right\}\times\]
\begin{equation}
\label{ep7.16}
[\hat{F}(z)e^{-\frac {3z^2} {2}}]
\phi(k)e^{kz+\frac {z^2} {2}-k^2}\;\frac {dk\;dz} {\sqrt{2}\;\pi}=
\end{equation}
\[\oint\limits_{{\Gamma}_k}\oint\limits_{{\Gamma}^{'}_z}
\left\{\Theta[\Im(k)]\Theta[\Im(z)]-
\Theta[-\Im(k)]\Theta[-\Im(z)]\right\}\times\]
\begin{equation}
\label{ep7.17}
\hat{F}(z)
\phi(k)e^{kz-z^2-k^2}\;\frac {dk\;dz} {\sqrt{2}\;\pi}=
\end{equation}
and then
\[<-2z\hat{F}(z)+{\hat{F}}^{'}(z),\phi(z)>=\]
\[\oint\limits_{{\Gamma}_k}\oint\limits_{{\Gamma}^{'}_z}
\left\{\Theta[\Im(k)]\Theta[\Im(z)]-
\Theta[-\Im(k)]\Theta[-\Im(z)]\right\}\times\]
\[[-2z\hat{F}(z)+{\hat{F}}^{'}(z)]
\phi(k)e^{kz-z^2-k^2}\;\frac {dk\;dz} {\sqrt{2}\;\pi}=\]
\[-\oint\limits_{{\Gamma}_k}\oint\limits_{{\Gamma}^{'}_z}
\left\{\Theta[\Im(k)]\Theta[\Im(z)]-
\Theta[-\Im(k)]\Theta[-\Im(z)]\right\}\times\]
\[\hat{F}(z)k
\phi(k)e^{kz-z^2-k^2}\;\frac {dk\;dz} {\sqrt{2}\;\pi}=\]
\begin{equation}
\label{ep7.18}
<\hat{F}(z),-z\phi(z)>
\end{equation}
If we define:
\begin{equation}
\label{ep7.19}
a=-z\;\;\;;\;\;\;a^+=-2z+\frac {d} {dz}
\end{equation}
we have
\begin{equation}
\label{ep7.20}
[a,a^+]=I
\end{equation}
The vacuum state annihilated by $a$ is $\delta(z)e^{z^2}$. The orthonormalized
states obtained by successive application of $a^+$ are:
\begin{equation}
\label{ep7.21}
{\hat{F}}_n(z)=2^{\frac {1} {4}}{\pi}^{\frac {1} {2}}
\frac {{\delta}^{(n)}(z)e^{z^2}} {\sqrt{n!}}
\end{equation}
On the real axis we have
\begin{equation}
\label{ep7.22}
<\hat{F}(z),\phi(z)>=\int\limits_{-\infty}^{\infty}\int\limits_{-\infty}^{\infty}
\overline{\hat{f}}(x)\phi(k)e^{kx-x^2-k^2}\;\frac {dx\;dk} {\sqrt{2}\;\pi}
\end{equation}
where $\overline{\hat{f}}(x)$is given by Dirac's formula:
\begin{equation}
\label{ep7.23}
\hat{F}(z)=\frac {1} {2\pi i}\int\limits_{-\infty}^{\infty}
\frac {\overline{\hat{f}}(x)} {x-z}\;dx
\end{equation}

\end{document}